\documentclass[aps,prl,twocolumn,amsmath,amssymb,showpacs,floatfix]{revtex4}
\usepackage{overpic}
\usepackage{amsmath}
\usepackage{amssymb}
 \usepackage{lmodern}
\usepackage{color}
\usepackage{graphicx}
\usepackage{booktabs}
\usepackage{xcolor,graphicx}
\usepackage{multirow}
\usepackage{upgreek}

\newcommand*{\ellp}{\ell_{\rm P}}
\newcommand*{\ellk}{\ell_{\rm K}}
\newcommand*{\weff}{w_{\rm eff}}
\renewcommand{\dh}{D_{\rm H}}
\newcommand*{\dw}{D_{\rm W}}
\newcommand*{\is}{I_{\rm s}}

\begin{document}
\title{Extension of nano-confined DNA: quantitative comparison between experiment and theory}

\author{V. Iarko$^1$, E. Werner$^1$, L. K. Nyberg$^2$, V. M\"uller$^2$, 
J. Fritzsche$^3$, T. Ambj\"ornsson$^4$, J. P. Beech$^5$, J. O. Tegenfeldt$^{5,6}$, 
K. Mehlig$^7$, F. Westerlund$^2$, B. Mehlig$^1$}
\affiliation{\mbox{}$^1$Department of Physics, University of Gothenburg, Sweden}
\affiliation{\mbox{}$^2$ Department of Biology and Biological Engineering, Chalmers University of Technology, Sweden}
\affiliation{\mbox{}$^3$ Department of Applied Physics, Chalmers University of Technology, Sweden}
\affiliation{\mbox{}$^4$Department of Astronomy and Theoretical Physics, Lund University, Sweden}
\affiliation{\mbox{}$^5$Department of Physics, Division of Solid State Physics, Lund University, Sweden}
\affiliation{\mbox{}$^6$NanoLund, Lund University, Sweden}
\affiliation{\mbox{}$^7$Department of Public Health and Community Medicine, University of Gothenburg, Sweden}

\begin{abstract}
The extension of DNA confined to nanochannels has been studied intensively and in detail. Yet quantitative comparisons between experiments and model calculations  are difficult because most theoretical predictions involve undetermined prefactors, and because the model parameters (contour length, Kuhn length, effective width) are difficult to compute reliably, leading to substantial uncertainties. Here we use a recent asymptotically exact theory for the DNA extension in the \lq extended de Gennes regime\rq{} that allows us to compare experimental results with theory. For this purpose we performed new experiments, measuring the mean DNA extension and its standard deviation while varying the channel geometry, dye intercalation ratio, and ionic buffer strength. The experimental results agree very well with theory at high ionic strengths, indicating that the model parameters are reliable. At low ionic strengths the agreement is less good. We discuss possible reasons. Our approach allows, in principle, to measure the Kuhn length and effective width of a single DNA molecule and more generally of semiflexible polymers in solution.
  \end{abstract}
\maketitle

Nano-confined DNA has recently been intensively studied
\cite{reisner2012,jones2013,gupta2014,Kho14,alizadehheidari2015c,reinhart2015} as a means of stretching the molecules in order to study local properties (e.g. DNA sequence \cite{lam2012,nilsson2014}). A fundamental question is how the physical properties of the DNA and the solution affect the extent to which the molecule is stretched by confinement. Experimentally this question has been investigated in detail, varying the confinement, the length of the DNA molecule, 
and the properties of the solution (see Ref.~\cite{reisner2012} for a review).

It is commonly assumed that DNA can be modeled as a semiflexible 
polymer with hard-core repulsive interactions \cite{grosberg1994,odijk2008,wang2011,tree2012,dai2013,muralidhar2014,muralidhar2014a}. 
Measurements \cite{hagerman1988,smith1992a,wang1997,bouchiat1999} and theoretical 
considerations \cite{hagerman1988,grosberg1994} indicate that 
a worm-like chain model may be a good approximation. 
A recent study \cite{gupta2014} compares experimental results for the
extension of confined $\lambda$-DNA 
to results of computer simulations of a self-avoiding discrete 
worm-like chain model, indicating that it may describe the experimental results well.

Yet quantitative comparison between experiments and theoretical model calculations has remained difficult for at least two reasons. First the model parameters 
(contour length, Kuhn length $\ellk$, and effective width $\weff$ of the semiflexible polymer) are difficult to determine reliably: there is substantial uncertainty
regarding the physical properties of DNA.
 Second, the model is hard to analyse theoretically.
But a recent asymptotically exact theory \cite{werner2014,werner2015} for the extension of a confined self-avoiding semiflexible 
polymer in the so-called \lq extended de Gennes regime\rq{} 
\cite{wang2011,dai2013,dai2014b} overcomes the second difficulty: it  makes precise predictions for the prefactors
and exponents defining scaling laws as a function of the physical parameters, Eqs.~(\ref{eq:edgPredictions}) below.
This opens the possibility to experimentally determine $\ellk$ and $\weff$ by measurements of confined DNA.
In this article we report on experimental results mapping out how
the extension of confined DNA in the extended de Gennes
regime depends on channel geometry, ionic strength of the solution, and upon
the amount of dye bound to the molecule.

At high ionic strengths we find very good agreement between experiment and theory 
using approximations for $\ellk$ and $\weff$ 
that are commonly employed~\cite{dobrynin2006,stigter1987,2007Reisner,2008Doyle},
and taking into account how the contour length depends on the amount of dye
molecules bound to the DNA. The comparison between experiment and theory is so precise that
it enables us to detect subtle alignment effects \cite{werner2012} at the border
of the extended de Gennes regime.  At low ionic strengths the agreement is not as good.
This may indicate that theoretical estimates of $\ellk$ and $\weff$ must be improved. 
We expect it is possible to experimentally precisely determine $\ellk$ and $\weff$ 
by extending the approach described in this article. 

{\em Extended de Gennes regime.} Consider a semiflexible polymer of contour length 
$L$, Kuhn length $\ellk$ \cite{grosberg1994} and excluded volume $v$ per Kuhn length 
segment. 
The excluded volume is often written in terms
of an effective width $\weff$, defined by the relation $v\equiv(\pi/2) \ellk^2\weff$. 
This expression for the excluded volume is based on Onsager's result \cite{onsager1949}
for the excluded volume of a cylinder of length $\ellk$ and diameter $\weff$, which in 
the limit $\ellk\gg \weff$ reduces to the expression above. 
We phrase our results in terms of the effective width $\weff$. This is customary but 
we stress that, strictly speaking, it is the excluded volume that determines the statistics of the polymer, and 
that even for DNA models with only hard-core repulsion between rod-like segments, the 
effective width does not equal the actual width of the rods, except in the limit 
$\ellk\gg \weff$. This is important to consider when evaluating the results of 
simulations.

The polymer is confined to a channel 
with 
cross section $\dw \times \dh$. The polymer exhibits different 
confinement regimes 
distinguished by different laws for 
the polymer extension 
in the channel direction \cite{werner2015}. 
The extended de Gennes regime is defined by the conditions \cite{werner2015}
\begin{equation}
\label{eq:edgConditions}
\ellk \ll \dh \ll \ellk^2/\weff\quad\mbox{and}\quad \dw^2 \ll \dh \ellk^2/\weff\,,
\end{equation}
where we assume that $\dw\ge \dh$.
For a square channel the corresponding conditions were 
previously derived 
and discussed in Refs.~\cite{odijk2008,wang2011,dai2013,dai2014b,werner2014}.
In regime (\ref{eq:edgConditions}) exact expressions for the mean $\mu$ and the variance $\sigma^2$ of the 
extension 
in the channel direction are known \cite{werner2014,werner2015}, provided
that the contour length is long enough \cite{werner2014}:
\begin{subequations}
\label{eq:edgPredictions}
\begin{align}\label{eq:mean}
 &\mu / L = 0.9338(84)\,[\ellk \weff/(\dw\dh)]^{1/3}\,,\\
\label{eq:variance}
 &\sigma / (L  \ellk)^{1/2} = 0.364(17)\,.
\end{align}
\end{subequations}
The errors quoted for the coefficients reflect strict bounds \cite{werner2014}
derived from the exact results of Ref.~\cite{vanderhofstad2003}.
Provided $L$ is known Eqs.~(\ref{eq:edgPredictions}) allow to infer $\ellk$ and $\weff$  
from measurements of nanoconfined DNA molecules. 
\begin{figure}
\includegraphics[width=0.95\columnwidth]{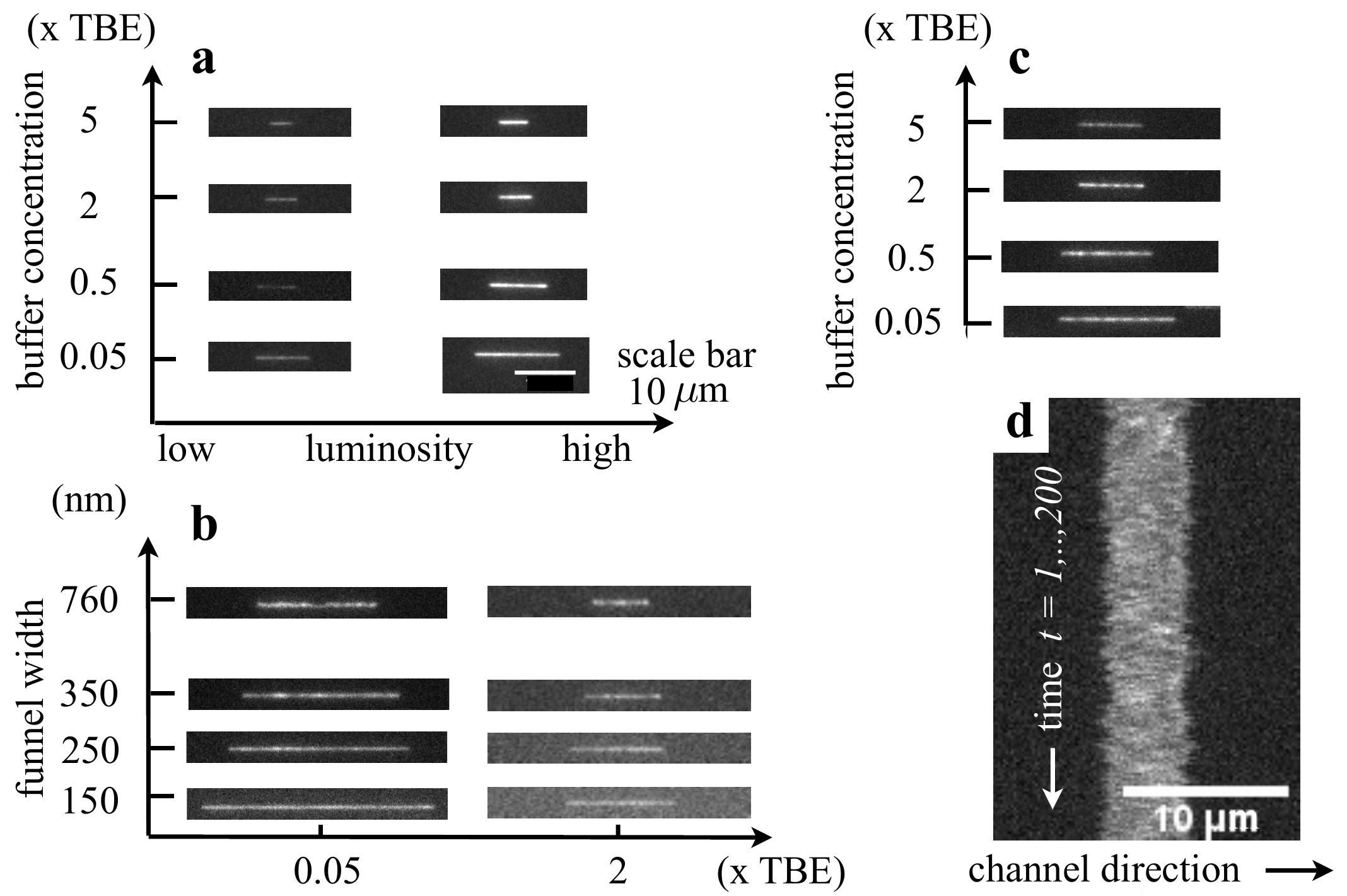}
\caption{\label{fig:summaryExperiments}
{\bf a} Experiment 1. 
$\lambda$-DNA in a $150$ nm $\times$ $108$ nm channel,
different buffer concentrations, and different luminosities
corresponding to different dye loadings. Shown 
are representative video frames
(scale bar applies to panels {\bf b},{\bf c} as well).
{\bf b} Experiment 2. 
T$4$-DNA in a nanofunnel in $0.05\times$ and $2\times$TBE 
solution, varying funnel width $\dw$ at constant $\dh=120$ nm. 
{\bf c} Experiment 3. 
T$4$-DNA in a $302$ nm $\times$ $300$ nm channel,
different buffer concentrations. 
{\bf d} Time trace of
the fluorescence intensity for $\lambda$-DNA in a 
$108\,{\rm  nm}\times 150\,{\rm  nm}$ channel in $5\times$TBE solution,  center-of-mass motion subtracted. 
}
\end{figure}

{\em Calculation of parameters}. 
We now discuss
how the parameters $L$, $\ellk$ and $\weff$ are commonly estimated for DNA,
and what the main uncertainties are. We first discuss bare DNA, before considering the 
effect of staining with fluorescent dye (YOYO-1).

The contour length of bare DNA is 0.34~nm
per base pair, with an uncertainty of about 0.01~nm, or 3\% \cite{sinden2012}.

DNA is commonly modeled as a worm-like chain, for which the Kuhn length is twice the
persistence length, $\ellk = 2\ellp$ \cite{grosberg1994}. 
The persistence length has been measured by a number of different techniques 
\cite{hagerman1988,smith1992a,B1997,bouchiat1999} yielding $\ellp\approx$ $45$--$50$~nm
at high ionic strength ($\is\gtrsim 10$~mM), and increasing at lower $\is$.
Following Refs.~\cite{2007Reisner,2008Doyle,nyberg2013,gupta2014} we use the empirical 
formula suggested in Ref.~\cite{dobrynin2006}
 $ \ellp \approx 46 {\rm nm}
+ {1.92 }\,\,{{\is^{-1/2}}}\, {\rm M}^{1/2}$ nm.
While this dependence of $\ellp$ upon $\is$ has a theoretical basis, the 
prefactors are known only from an empirical fit with an uncertainty of about 10\% 
(Fig.~3 in Ref.~\cite{dobrynin2006}). The resulting values of $\ellk$
are given in Table~\ref{tab:1}, the calculation of  $\is$ is described
in the Supplemental Material \cite{supp}.

Stigter \cite{stigter1987} computed the excluded volume
between two long, strongly charged cylinders in NaCl solution, and applied this 
calculation to DNA to obtain an estimate of $\weff$.
Linear interpolation on a doubly logarithmic scale of the effective widths given in 
Table~1 of Ref.~\cite{stigter1987} 
yields the values tabulated in Table~\ref{tab:1}.
There are many sources of uncertainty when applying this theory to our system. 
Stigter's calculation \cite{stigter1987} for $\weff$ assumes that the 
Kuhn length segments can be approximated by infinitely long cylinders with an 
intrinsic width of 1.2 nm, and that the effective line
charge of DNA is given by $0.73e^-$ per phosphate group.
The approximation of infinite cylinders is problematic when the Kuhn length and 
the effective width are of the same order, i.e. for low ionic strengths. According
to Stigter, the value 1.2~nm has an uncertainty of about 20\%, leading to an 
uncertainty for the effective 
width of 5-10\% \cite{stigter1977}, with a larger effect at large ionic strengths.
The effective line charge estimate is based on measurements in NaCl-solutions, which 
do not generalise to other ions \cite{schellman1977}. 
\begin{table}[t]
  \centering
  \begin{tabular}{lllll}
  \hline\hline
 &\text{$0.05\times$TBE} & \text{$0.5\times$TBE}  & \text{$2\times$TBE} & 
 \text{$5\times$TBE} \\
 \hline
 \text{$\is$[mM]}\phantom{xxx}    & 3.81 & 24.9 & 78.4 & 178.0 \\
 \text{$\ellk$[nm]}  & 154.4 & 116.5 & 105.9 & 101.2\\
 \text{$\weff$[nm]}    &  26 & 10 & 6.2 & 4.6 \\
  \hline
  \hline
  \end{tabular}
  \caption{\label{tab:1} Numerical values for the ionic buffer strength $\is$, the Kuhn length $\ellk$, and the effective width $\weff$ (see text). }
\end{table}

The DNA contour length is expected to increase in proportion to the amount of dye 
bound.
We assume that dye intercalation extends the bare 
contour length of the DNA molecule by $0.44$~nm per dye molecule~\cite{gupta2014}.
Estimates of this 
number range from about 0.4~nm to 0.5~nm, with large uncertainties in the individual 
estimates \cite{johansen1998,gunther2010}, the uncertainty is at least 10\%. At a dye
loading of 1 molecule per 10 base pairs, this corresponds to an uncertainty in the
contour length of $\approx 2\%$.
There is no consensus regarding how intercalating dye molecules affect the parameters 
$\ellk$ and $\weff$. 
Ref.~\cite{murade2010} finds that the Kuhn length decreases with increasing dye load, 
whereas Ref.~\cite{gunther2010} finds no dependence. Since the dye molecules are 
positively charged, 
the effective width might decrease with dye load, but the magnitude of this
effect is not known. Lacking a better estimate, it is commonly assumed that YOYO-
binding does not affect these parameters \cite{reisner2005,2008Doyle,gupta2014}.
In summary, while the contour length of DNA is known to a rather high accuracy, there 
is substantial uncertainty regarding the parameters $\ellk$ and $\weff$. 

{\em Experimental method.}
The experimental data are obtained measuring
the extension of single DNA molecules in nanochannels
under different conditions (Fig.~\ref{fig:summaryExperiments}{\bf a} to {\bf c}). 
We use linear $\lambda$- and T$4$GT$7$-DNA (T$4$-DNA for short)
with definite contour lengths of 
48502 and 165647 base pairs, respectively (this ensures
that $L$ is large enough in our experiments, it
exceeds  $(\dw^{2}\dh^{2}\ellk/\weff^{2})^{1/3}$
by an order of magnitude \cite{werner2015}).
The molecules are stained with YOYO dye and suspended inside a channel in a TBE (Tris-Borate-EDTA) buffer. 

The first experiment is a re-analysis of data presented in Ref.~\cite{nyberg2013}. 
In this experiment (Fig.~\ref{fig:summaryExperiments}{\bf a}), $\lambda$-DNA is 
inserted into a nanochannel of height $\dh = 150$~nm and  width $\dw = 108$~nm.
We discuss the uncertainty in
the channel dimensions in the Supplemental Material~\cite{supp}. DNA extensions
are measured at different buffer conditions 
($0.05\times$, $0.5\times$, $2\times$ and $5\times$TBE), 
and at  different dye loads.
To estimate the dye load of a molecule we assume that it is proportional to the 
luminosity, and
that the largest observed luminosity corresponds 
to full intercalation (one dye molecule per four base pairs). 
In this way we obtain an estimate for the amount of dye bound to the molecule by
linear interpolation.

In experiment 2 (Fig.~\ref{fig:summaryExperiments}{\bf b}), T$4$-DNA is inserted into a
nanofunnel, with 
fixed height $\dh = 120$~nm and  gradually changing width from $\dw=92$~nm 
to $\dw=815$~nm over a length of 500~$\upmu$m. 
These experiments are at two different buffer concentrations 
($0.05\times$ and $2\times$TBE).

In experiment 3 (Fig.~\ref{fig:summaryExperiments}{\bf c}), T$4$-DNA is 
inserted into a channel with $\dw=302$~nm, $\dh=300$~nm. The buffer concentration is varied
($0.05\times$, $0.5\times$, $2\times$, and $5\times$TBE). In experiments 2 and 3, the average
dye load at 0.05$\times$TBE is approximately 1 dye molecule 
per 10 base pairs. Assuming that the dye load is proportional to luminosity we
estimate the dye load in experiment 2 at $2\times$TBE to 1 dye molecule per 45 base pairs,
and 1 per 12, 16, 28 base pairs at $0.5\times$, $2\times$, and 5$\times$TBE in experiment 3.

For each molecule $200$ frames are recorded.
Fig.~\ref{fig:summaryExperiments}{\bf d} shows an example of a fluorescence-intensity 
trace (\lq kymograph\rq{}) obtained in this way. Each row in the kymograph shows the
fluorescence intensity in a given frame
averaged over the channel cross section. 
Bright regions correspond to high intensity indicating
where the DNA molecule is located. The extension of the molecule along the channel is 
simply the width of the bright region.

For a given set of parameters we estimate the mean and standard 
deviation of the extension by a linear mixed model that takes into account the fact 
that the measured extensions are correlated in time.
Details concerning the experimental method and the
data analysis are given in the Supplemental Material~\cite{supp}.

\begin{figure*}[t]
\begin{overpic}[width=\textwidth]{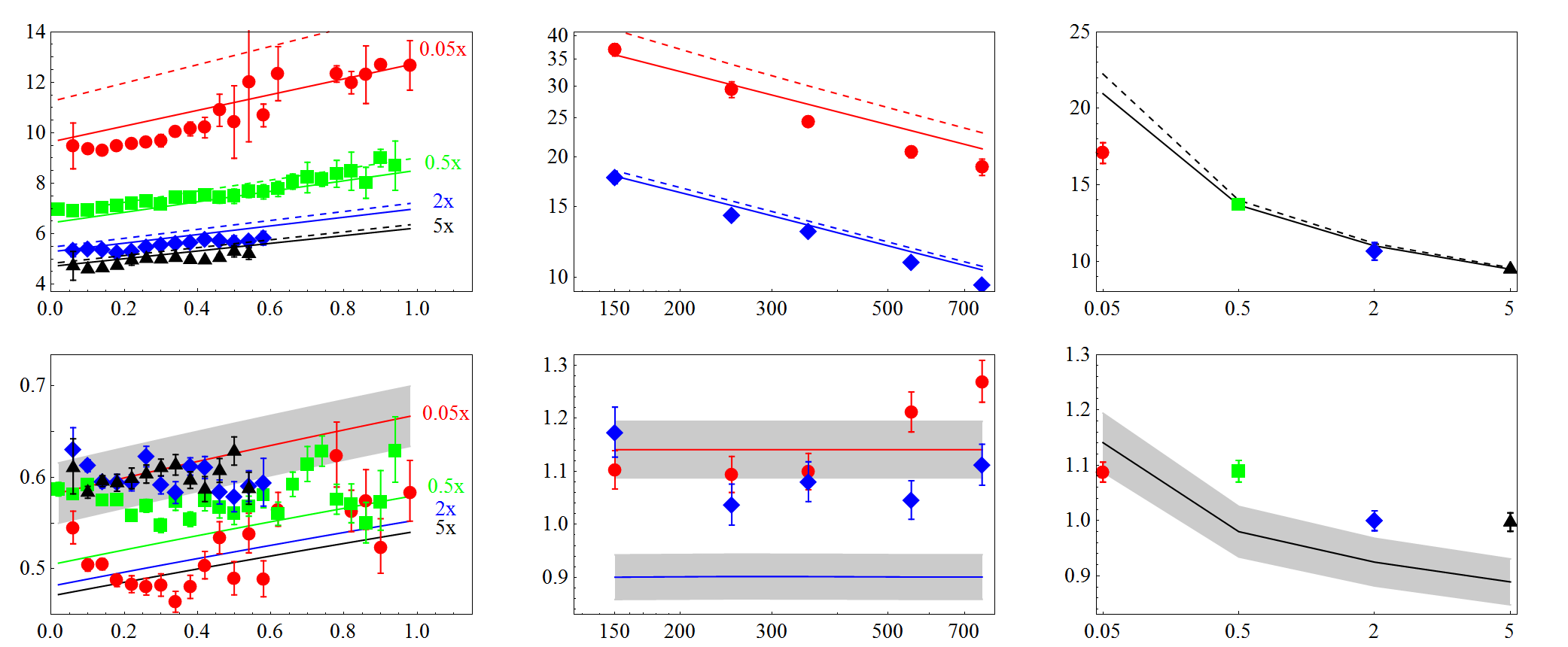}
\put(3.28,37.7){\colorbox{white}{\bf{a}}}
\put(3.4,16.6){\colorbox{white}{\bf{b}}}
\put(37.2,37.8){{\bf{c}}}
\put(36.9,16.6){\colorbox{white}{\bf{d}}}
\put(70.3,37.5){\colorbox{white}{\bf{e}}}
\put(70.3,16.4){\colorbox{white}{\bf{f}}}
\put(-1,27){\rotatebox{90}{\colorbox{white}{$\mu$, $\upmu$m }}}
\put(-1,7){\rotatebox{90}{\colorbox{white}{$\sigma$, $\upmu$m }}}
\put(8,-2){\rotatebox{0}{\colorbox{white}{Relative luminosity}}}
\put(46,-2){\rotatebox{0}{\colorbox{white}{$\dw$, nm }}}
\put(71,-2){\rotatebox{0}{\colorbox{white}{Buffer concentration ($\times$TBE)}}}
\end{overpic}
\mbox{}\\[-2mm]
\caption{\label{fig:experimentalResults}
Experimental results for 
$0.05\times$TBE (red $\circ$), 
$0.5\times$TBE (green $\Box$), $2\times$TBE (blue $\Diamond$),
and $5\times$TBE (black $\bigtriangleup$)
{\bf a}, {\bf b} Experiment 1. Mean and standard deviation of the extension of 
$\lambda$-DNA
in a narrow nanochannel, as a function of relative luminosity. 
Theory [Eq.~(\ref{eq:edgPredictions})], solid lines.
The rigorous bounds on the prefactor in Eq.~(\ref{eq:variance}) are indicated as a 
shaded region for $0.05\times$TBE, they are of the same order for the other cases. 
The corresponding uncertainty for the extension is much smaller and not shown. 
The dashed line shows theory corrected for wall repulsion (see text).
{\bf c}, {\bf d} Experiment 2. Same, but for T$4$-DNA
in  a nanofunnel with varying width $\dw$. Note that panel {\bf c} is a 
log-log~plot.
{\bf e}, {\bf f} Experiment 3. Same, but for T$4$-DNA
in a wider square nanochannel, as
a function of buffer concentration ($x\times$TBE).
Error bars correspond to 95\% confidence intervals from the statistical analysis, 
the experimental uncertainty is not taken into account.
}
\end{figure*}
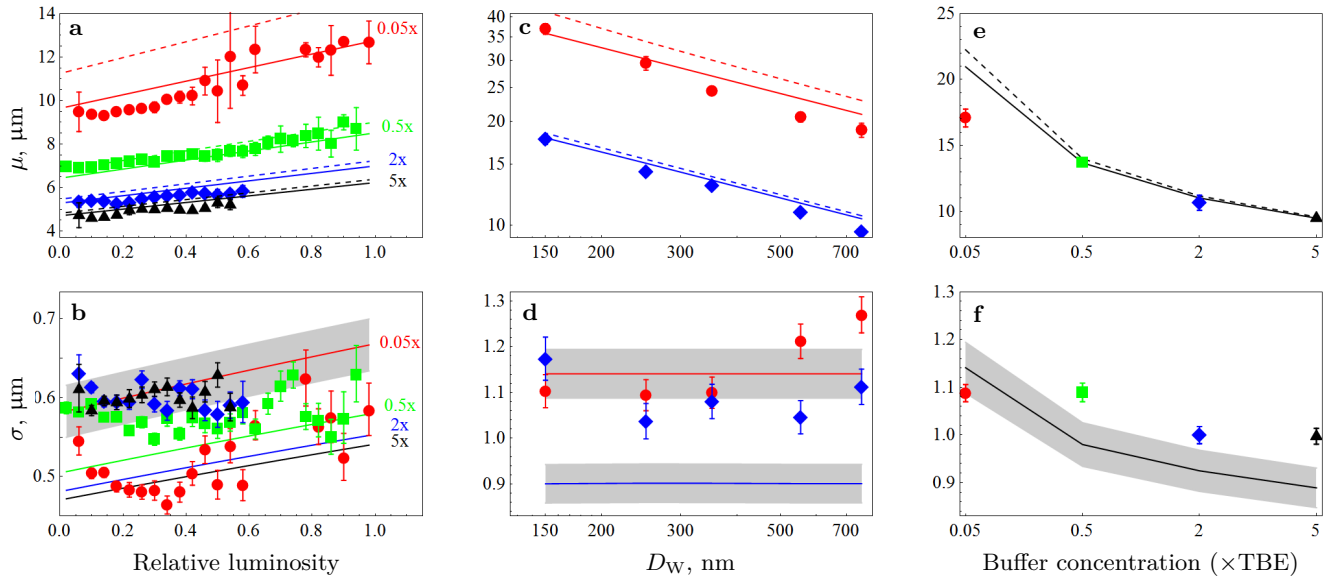	

{\em Results.} 
Our results are shown in 
Fig.~\ref{fig:experimentalResults}.
We plot two theoretical curves. The solid curve uses the actual
channel size $\dh \times \dw$. The dashed curve compensates for the
repulsive interaction with the negatively charged walls \cite{reisner2012}, by 
using an `effective channel size' $(\dh- \delta) \times (\dw - \delta)$. 
We take $\delta=\weff$,
but it is not 
known how accurate this estimate is.
Since the standard deviation is independent of channel size in the extended de Gennes
regime, the compensation does not affect this comparison.

The results of experiment 1 are shown in panels {\bf a} and {\bf b}. At low relative 
luminosity (small dye-to-basepair ratio) the average extension is well described by 
Eq.~(\ref{eq:mean}).
For the standard deviation there are larger differences between experiment
and Eq.~(\ref{eq:variance}). Possible reasons are discussed below.

We turn now to the effect of increasing the dye-to-basepair ratio of the DNA.
The theoretical lines are calculated under the assumption that each dye molecule increases the contour 
length by 0.44~nm but leaves the Kuhn length and the effective width unchanged. This
yields estimates of the mean and standard deviation that overestimate the observables at high
ionic strengths and high dye loads. A simple explanation would be that the persistence 
length decreases slightly with increasing dye load, in agreement with Ref.~\cite{murade2010} though
not with Ref.~\cite{gunther2010}. Note that since experiments 2 and 3 were performed at 
low dye-to-basepair ratios, such a decrease would not significantly influence the interpretations
of these experiments.

The results of experiment 2 are shown in panels {\bf c}, {\bf d}. Again, the 
experimental results are in qualitative agreement with the theoretical predictions. 
We see that the average extension agrees well with the theoretical prediction. 
However, the model predictions underestimate the 
standard deviation for the larger ionic strength, and for the largest channel at the 
lower ionic strength.

It is important to note that for experiment 2 we do not expect perfect 
agreement with Eqs.~(\ref{eq:edgPredictions}), since the 
condition $\dh \gg \ellk$ is not satisfied, or only weakly satisfied. However, as long 
as $\dw\gg \ellk$, the violation of the condition for $\dh$ only affects the 
prefactors but not the power of $\dw$ in Eqs.~(\ref{eq:edgPredictions}). 
This follows from the 
fact that a mapping to a one-dimensional model is possible also when 
$\dh \approx \ellk$ 
\cite{werner2015}. In accordance with this prediction, the data points satisfying
$\dw\gg\ellk$ in panel {\bf c} obey the scaling $\mu\propto \dw^{-1/3}$ of 
(Eq.~\ref{eq:mean}). Similarly the data points at 2$\times$TBE which satisfy 
$\dw\gg\ellk$ show a 
variance that is approximately independent of $\dw$, in agreement with 
Eq.~\ref{eq:variance}, Fig.~\ref{fig:experimentalResults}{\bf d}.
For the two rightmost data points at 0.05$\times$TBE, 
the condition $\dw^2 \ll \dh \ellk^2/\weff$ is violated. At this point the variance 
is expected to start increasing as $\dw$ increases further \cite{werner2015}, in 
perfect agreement with what is observed.

For square channels simulations \cite{wang2011,dai2014b,muralidhar2014a} show that the 
mean extension 
increases more rapidly with decreasing $D=\dh=\dw$ than Eq.~(\ref{eq:edgPredictions}a) predicts, when $D \approx \ellk$. The reason is that there is a 
tendency for the DNA molecule to align with the wall, and that the presence of the 
walls makes it more difficult for the molecule to change direction in the channel, 
forming a `hairpin' \cite{werner2012,muralidhar2014a}. This can explain why
the average extension appears to increase slightly faster with 
decreasing width than Eq.~(\ref{eq:edgPredictions}a) predicts, for the 
leftmost points in panel~{\bf c}. Such a trend has also been
observed in previous measurements in rectangular channels 
\cite{persson2009,werner2012}.

Now consider the standard deviation. The alignment and correlation effects mentioned above 
cause $\sigma$ to be overestimated \cite{werner2012}. But when $\mu$ approaches the maximal extension $L$ then
fluctuations are suppressed \cite{muralidhar2014a}. These two effects could explain 
why, in experiments 1 and 2, $\sigma$ is larger than predicted by theory at high ionic 
strengths but 
smaller for low ionic strengths and small channel sizes. It must also be noted that
the standard deviation is difficult to estimate precisely, as it is not very much 
larger than the pixel size in the image (159~nm), 
and may depend on the assumptions entering into the statistical analysis (see Supplemental Material \cite{supp}).
An additional source of uncertainty specific to experiment 2 is that $\dw$ changes 
over the span of the molecule. For the most extended condition 
($\mu= 37\,\upmu$m), the channel
width at either end of the molecule differs by approximately 25~nm from the stated 
width, measured at the center of the molecule. 

The results of experiment 3 are shown in panels {\bf e}, {\bf f}. Here
Eq.~(\ref{eq:edgConditions}) is well satisfied. 
Simulations indicate
\cite{muralidhar2014a,dai2014b} that the alignment effects
discussed above have little influence on $\mu$, in square channels 
with $D\approx 3\ellk$. Equally sensitive simulation results for $\sigma$  have not 
been published, but simulations of the alignment effect
\cite{werner2012,muralidhar2014a} indicate that Eq.~(\ref{eq:variance}) underestimates
$\sigma$ by approximately 10\%. 
We find that for the three largest ionic strengths, measurements are in excellent agreement with theory. The mean extension (panel {\bf f}) 
agrees very well with the theoretical prediction of Eq.~(\ref{eq:mean}), and 
Eq.(\ref{eq:variance}) underestimates the standard deviation (panel {\bf f}) by 
about 10\%, just as the measurements of alignment effects would suggest. 

Intriguingly, the relation between measurements and predictions is different at 
0.05$\times$TBE than 
at high ionic strengths. Both mean and standard deviation are smaller than expected. This is 
particularly surprising considering that alignment effects should be even stronger at 
low ionic strength (where $\ellk$ is larger).
The discrepancy might indicate that the standard model does not describe the 
physical parameters well at such low ionic strengths, possibly because the high 
relative concentration of BME significantly 
changes the buffer conditions, lowering the pH from $\approx 8.5$ to $\approx 7.5$.
But we note that it may be hard to ensure uniform dye coverage under these conditions 
\cite{nyberg2013}.  Also, BME is consumed as the experiment proceeds. This may change the ionic strength at small buffer
concentrations, an effect our calculations do not include.

{\em Conclusions.} We have compared measurements of the extension of confined
DNA to asymptotically exact predictions.
First, we find very good agreement between 
experiments and theoretical predictions at high ionic strengths. 
A possible cause for deviations at low ionic strengths is that common
estimates for $\weff$ and $\ellk$ of stained DNA are too imprecise. 
Second, by measuring longer time series and more molecules in wider channels  we
expect to be able to precisely determine how $\ellk$ and $\weff$ depend on the ionic 
strength of the solution. It may even be possible to determine $\ellk$ and $\weff$ for single molecules.
We note  that $\weff$ is an effective parameter
defined in terms of the excluded volume per Kuhn length, even for
rigid rods $\weff$ is identical to the actual width only when $\ellk\gg \weff$.
Finally the approach described here could be used to investigate
DNA-wall interactions, a question about which little is known, and that is hard 
to describe theoretically.

{\em Acknowledgements.}
This work was made possible through support from the Swedish Research Council (BM), the G\"oran Gustafsson Foundation for Research in Natural Sciences and Medicine (BM), Chalmers Area of Advance (FW). We thank Charleston Noble and Erik Lagerstedt for  helpful discussions. 

After submission of this manuscript a paper \cite{gupta2015} appeared online, which
also compares experimental measurements on channel-confined DNA to the theoretical 
predictions of Ref.~\cite{werner2014}.

\end{document}


\title{{\rm Supplemental material for}\\Extension of nano-confined DNA: quantitative comparison between experiment and theory}

\author{V. Iarko$^1$, E. Werner$^1$, L. K. Nyberg$^2$, V. M\"uller$^2$, 
J. Fritzsche$^3$, T. Ambj\"ornsson$^4$, J. P. Beech$^5$, J. O. Tegenfeldt$^{5,6}$, 
K. Mehlig$^7$, F. Westerlund$^2$, B. Mehlig$^1$}
\affiliation{\mbox{}$^1$Department of Physics, University of Gothenburg, Sweden}
\affiliation{\mbox{}$^2$ Department of Biology and Biological Engineering, Chalmers University of Technology, Sweden}
\affiliation{\mbox{}$^3$ Department of Applied Physics, Chalmers University of Technology, Sweden}
\affiliation{\mbox{}$^4$Department of Astronomy and Theoretical Physics, Lund University, Sweden}
\affiliation{\mbox{}$^5$Department of Physics, Division of Solid State Physics, Lund University, Sweden}
\affiliation{\mbox{}$^6$NanoLund, Lund University, Sweden}
\affiliation{\mbox{}$^7$Department of Public Health and Community Medicine, University of Gothenburg, Sweden}
\maketitle

\section{Experimental procedure}        
{\em Channel manufacture.}
The channels were manufactured as described in Ref.~\cite{persson2010}.
The nanochannels used in experiment 1 have a depth of $\dhh=150$~nm.
The channel cross-section is not perfectly rectangular, but rather trapezoidal, with 
a width at the top (bottom) of the channel of 130~nm (87~nm). We assume that these 
channels can be approximated by rectangular channels with a width $\dw=108$~nm, which 
is the arithmetic mean of the measurements at the top and bottom of the channel.
The funnels used in experiment 2 have a depth of 
approximately $\dhh = 120$ nm, and $\dw$ increases from $92$~nm 
(top: 111~nm, bottom: 73~nm) at the narrow end to $815$~nm 
(top: 830~nm, bottom: 800~nm) at the wide end -- over a length of 500~$\upmu$m. 
Finally, the channels used in experiment 3 have a depth $\dhh=300$~nm, and a width
$\dw=302$~nm (top: 330~nm, bottom: 275~nm).
The dimensions are measured before the channel is closed by the lid.
Since the bonding process changes the depth of the channel, our values for the depth
of the channels have a relatively large uncertainty of about 10~nm. 

{\em Buffer preparation.} 
The buffers were obtained by
diluting $10\times$TBE tablets from Medicago to the desired concentration.
$1\times$TBE contains $0.089$ M Tris, $0.089$ M Borate and $0.0020$ M EDTA.
Right before the experiments, 3\% BME (3 $\mu$L of BME to 97 $\mu$L of
sample solution) was added to prevent photo-nicking of the DNA. In addition to this the micro- and nanochannels in the chip
were flushed with the right buffer concentration, containing 3\% BME as well, to keep
the same environment in the chip as in the sample solution.

{\em Dye intercalation.}
Since intercalation of YOYO-dye molecules extends the contour length of the DNA
it is important to estimate the dye load as accurately as possible.
In experiment 1, the dye load was not equilibrated between molecules, resulting in 
heterogeneous staining \cite{nyberg2013}.
In experiments 2 and 3 the aim was instead to achieve homogeneous staining.
To this end we first mixed the
sample in high ionic strength ($5\times$TBE), let it rest for at least 20 min, and 
then diluted to the desired ionic strength \cite{nyberg2013}.

{\em DNA insertion.} 
The nanofluidic chips consist of four loading wells connected two and two by 
microchannels. The microchannels are in turn connected by many nanochannels. Sample 
solution was inserted into one of the loading wells. Then the DNA was carried to the 
nanochannels by pressure-driven flow (N2 gas for T4 experiments and air for lambda 
experiments). The DNA was forced into the nanochannels by applying pressure over two 
wells that are connected by a microchannel. Once inserted in the channel, the pressure
was switched off. Before imaging the DNA was allowed to relax for about 30 seconds to 
reach its equilibrium extension.

{\em Video recording.}
Videos were recorded using a Photometrics Evolve\textsuperscript{TM} EMCCD camera. 
The image pixel size is 159.2 nm.
For all measurements, the exposure time was 100 ms per photograph, but the 
delay between frames differed between experiments.
For experiment 1, the delay between frames was 84 ms, yielding a frame-rate of 
184~ms/frame.
For experiments 2-3 with T4-DNA, the correlation time is significantly longer, 
so to minimise problems with photo bleaching and nicking, the frame-rate was decreased.
For experiment 2, the frame-rate was 0.5~s/frame for the measurements
at $0.05\times$ TBE, and 2~s/frame for the ones at $2\times$ TBE.
For experiment 3, the frame-rate was 1~s/frame for all measurements.

{\em Number of molecules.}
In experiment 1, 2388 molecules were analysed in total. Since the relative 
intensities are distributed unevenly, so are the number of molecules in each bin.
This leads to very different error estimates for different bins in the estimation of 
the mean and standard deviation. We excluded bins with 2 molecules or less from the 
analysis. Further 3 molecules were excluded because the extension could not be 
succesfully extracted from all frames.
In experiment 2, 9-10 molecules were analysed per data point under 
0.05$\times$TBE, and 7 molecules under 2$\times$TBE. 
In experiment 3, 30 to 35 molecules were analysed per measured data point. 
One outlier at 0.5x TBE was removed, as its extension changed abruptly halfway through
the data series.

\section{Data analysis}
{\em Extraction of DNA extension from kymographs.}
Consider a raw kymograph of the form provided in the main text (Fig.~1d).
Dark regions in this image correspond to background, the bright regions
correspond to fluorescence from DNA. The fluorescence intensities
measured for both regions are
subject to noise. This makes it difficult to reliably identify the end points of the
bright regions in an automated fashion.
In previous studies this was achieved by fitting the difference of
two sigmoidal functions \cite{reisner2012} to each time frame. This method was used 
for analysing experiment 1 \cite{nyberg2013}. 
In experiments 2-3 we used a computationally faster method that yields
very similar results for the mean, and agrees to within about 5\% for the
standard deviation of the extension. 
The new method for detecting end positions relies on the 
assumption that the
fluorescence intensities from DNA and from the background assume consistent and easily 
differentiated values, and proceeds as follows. First, each frame
in the raw kymograph is smoothed using a moving 
average (we use an averaging window that is five pixels wide).
Second, each frame is then segmented into binary low- and high-intensity regions
using Otsu's method \cite{otsu1979,liao2001}. Third, we even out 
gaps between high-intensity regions that are equal to or shorter than five pixels.
Fourth, the largest connected high-intensity component is found, and its edges 
are identified. The distance between these edges is the extension of the DNA molecule.

{\em Statistical analysis.}
The experimental data consist of repeated measurements of the extension of individual 
DNA molecules at given values of parameters and are modeled using a 
random-coefficient model, a variation of a linear-regression model
adapted to correlated data \cite{Sni99}.
Let $y_{ij}$ denote the extension of molecule $i$ ($i=1,\dots, n$)
at time $t_j$ ($j=1,\dots, 200$).
The extension is modeled as a linear function of time as
$y_{ij} = \mu  + \beta t_j + u_{i} + e_{ij}$,
where $\mu$ and $\beta$  denote the overall mean value and slope.
To account for the within-molecule correlation of data points the error term is split 
into two uncorrelated components, a molecule-specific correction to the mean, $u_i$, 
and an error term $e_{ij}$ that measures the deviation of a
data point from the molecule-specific regression line, $\mu  + \beta t_j + u_{i}$. 
Both $u_i$ and $e_{ij}$ are assumed to be
independently normally distributed with mean zero and variance $r^2$ and $\sigma^2$, 
respectively. The model
describes the mean extension of molecules at a particular set of external parameters 
and the linear time dependence allows to include small unmeasured changes of external 
conditions, for instance drift into wider or narrower channel regions, or change of 
buffer concentration. The random intercept adjusts for the initial conditions at the
start 
of measurement of each time series. A random intercept model is adequate as each set 
of molecules can be viewed as a random sample representative for all molecules at this 
particular set of external parameters. On average, the covariance of two arbitrary 
data points within molecules is equal to the variance of intercepts
between molecules. The ratio of variances, $r^2 / (r^2 + \sigma^2)$ is both a measure 
for the proportion of
variation that is explained by differences between molecules, and the correlation of 
data points within a typical time series.
By isolating the molecule-specific random variation due to unknown fluctuations 
of external parameters we are able to obtain an improved value for the residual 
variance of the extension ($\sigma^2$)
that can be compared with the theory. Although the time trend was weak in most cases, 
we evaluated the mean value $\mu$ at $t = 100$, i.e. we replaced $t_j$ with 
$(t_j - 100)$ in the above model. Results
are given in terms of point estimates together with a 95\% confidence 
interval (95\% CI). The statistical analyses were performed by using 
SAS (version 9.4; SAS Institute, Cary, NC).

\section{Calculation of ionic strength}
The parameters $w$ and $\ellk$ of the DNA molecule depend on the
ionic strength of the surrounding solution. The ionic strength is defined as
\begin{align}
\label{eq:ionicStrength_general}
I_s=\frac{1}{2}\sum_i c_i z_i^2.
\end{align}
Here, the sum runs over all ions in the solution, $c_i$ is the
concentration of ion species $i$, and $z_i$ its valence.
To compute
the ionic strength, we must calculate the equilibrium concentrations $c_i$ of all
ions. Since there is some confusion in the literature (discussed below) about the
ionic strength of TBE buffer, we document our calculation in some detail.

We follow the method outlined in Ref.~\cite[Section~12-2]{harris2010}.
Denote by $C[X]$ the total amount of all species of a substance $X$, in neutral and
ionic form.
At $N\times$TBE, the solution contains $C[{\rm T}]=N\times0.089$ M Tris,
$C[{\rm Bo}]=N\times0.089$ M Borate,
$C[{\rm E}]=N\times0.0020$ M EDTA, and $C[{\rm \upbeta}]=0.429$~M BME. 
The dissociation and association constants
that determine the chemical equilibrium are
Tris: p$K_b=5.94$;
Borate: p$K_a=9.24$,
EDTA: p$K_a=\{1.99,2.67,6.16,10.26\}$;
BME: p$K_a=9.6$.
At the resulting pH-value of approximately
8.5, most of the EDTA is triply ionised, and we can safely ignore the minute
concentrations of neutral and singly ionised EDTA. We denote the molar concentration
of a species by [X], and its activity coefficient by $\gamma_{\rm X}$. The activity
coefficients are assumed to be given by the Davies equation, as stated in
Ref.~\cite{harris2010} (the empirical prefactors of this equation differ between
sources. For our ionic strength calculations, the difference between
different formulations makes a difference of two percent at most).
The system of equations that must be solved is
\begin{align}
\label{eq:firstMAEquation}
\massActionEquation{ TH^+}{OH^-}{T}{5.94} \\
\massActionEquation{ Bo^-}{H^+}{HBo}{9.24} \\
\massActionEquation{ HE^{3-}}{H^+}{H_2E^{2-}}{6.16} \\
\massActionEquation{ E^{4-}}{H^+}{HE^{3-}}{10.26} \\
\massActionEquation{ \upbeta^-}{H^+}{H\upbeta}{9.6} \label{eq:lastMAEquation}\\
[{\rm H^+}]\gamma_{\rm H^+}[{\rm OH^-}]\gamma_{\rm OH^-} &= 10^{-14.0} \label{eq:firstCEquation}\\
[{\rm T}] + [{\rm TH^+}] &= C[{\rm T}] \\
[{\rm HBo}] + [{\rm Bo^-}] &= C[{\rm Bo}] \\
[{\rm H_2E^{2-}}] + [{\rm HE^{3-}}] + [{\rm E^{4-}}] &= C[{\rm E}] \\
[{\rm H\upbeta}] + [{\rm \upbeta^-}] &= C[{\rm \upbeta}] \label{eq:lastCEquation}\\
[{\rm TH^+}] + [{\rm H^+}] &= \label{eq:chargeBalance}\\
[{\rm OH^-}] + [{\rm Bo^-}] + 2[{\rm H_2E^{2-}}] + 3[{\rm HE^{3-}}] &+ 4[{\rm E^{4-}}] + [{\rm \upbeta^-}] \nonumber
\end{align}
Eqs.~(\ref{eq:firstMAEquation})-(\ref{eq:lastMAEquation}) are the equilibrium conditions from the law of mass action, 
Eqs.~(\ref{eq:firstCEquation})-(\ref{eq:lastCEquation}) ensure that the total concentration of a substance is conserved, and 
Eq.~(\ref{eq:chargeBalance}) ensures charge neutrality.
The activity coefficients $\gamma_{\rm X}$ depend on the ionic strength $I_s$ 
according to the Davies equation \cite{harris2010}, 
\begin{align}
\label{eq:davies}
\log_{10}\gamma_{\rm X} &=-0.51 z_{\rm X}^2 \left(\frac{\sqrt{I_s}}{1+\sqrt{I_s}} - 0.3 I_s\right).
\end{align}
Here, $z_{\rm X}$ is the valence of species $X$, and $I_s$ is measured in units of M.
Finally, the ionic strength itself is given according to Eq.~\ref{eq:ionicStrength_general} by
\begin{align}
\label{eq:ionicStrength_specific}
I_s &= 1/2\bigg([{\rm TH^+}] + [{\rm H^+}] +[{\rm OH^-}] +[{\rm Bo^-}] + \\
 &4[{\rm H_2E^{2-}}] + 9[{\rm HE^{3-}}] + 16[{\rm E^{4-}}] + [{\rm \upbeta^-}]\bigg). \nonumber
\end{align}
We solve the system of 
equations~(\ref{eq:firstMAEquation})-(\ref{eq:ionicStrength_specific}) iteratively. 
Starting from the initial guess $\gamma_{\rm X}=1$ for all species, 
Eqs.~(\ref{eq:firstMAEquation})-(\ref{eq:chargeBalance}) are solved with the
Mathematica routine {\tt FindInstance}. Plugging the resulting concentrations into
Eqs.~(\ref{eq:davies})-(\ref{eq:ionicStrength_specific}) yields new values for 
$\gamma_{\rm X}$, which are then used in
Eqs.~(\ref{eq:firstMAEquation})-(\ref{eq:chargeBalance}). This process was repeated 
until the ionic strength converged.

We tested our calculations by comparing with Table~1 of Ref.~\cite{2008Doyle}. We
reproduce the reported ionic strengths for 0\% BME and 0.5\% BME to within one
percent.

\vfill